\documentclass{emulateapj}

\slugcomment{Submitted to ApJ Letters}

\shorttitle{The Warp of the Galaxy}
\shortauthors{Weinberg \& Blitz}

\begin{document}

\title{A Magellanic Origin for the Warp of the Galaxy}
\author{Martin D. Weinberg}
\affil{Astronomy Department, University of Massachusetts
    Amherst, MA 01003}
\and
\author{Leo Blitz}
\affil{Astronomy Department, University of California,
    Berkeley, CA 94720}
\email{weinberg@astro.umass.edu; blitz@berkeley.edu}

\begin{abstract}
We show that a Magellanic Cloud origin for the warp of the Milky Way
can explain most quantitative features of the outer HI layer recently
identified by Levine, Blitz \& Heiles (2005). We construct a model
similar to that of Weinberg (1998) that produces distortions in the
dark matter halo, and we calculate the combined effect of these
dark-halo distortions and the direct tidal forcing by the Magellanic
Clouds on the disk warp in the linear regime. The interaction of the
dark matter halo with the disk and resonances between the orbit of the
Clouds and the disk account for the large amplitudes observed for the
vertical $m=0,1,2$ harmonics.  The observations lead to six
constraints on warp forcing mechanisms and our model reasonably
approximates all six.  The disk is shown to be very dynamic,
constantly changing its shape as the Clouds proceed along their
orbit. We discuss the challenges to MOND placed by the observations.
\end{abstract}


\section{Introduction}

The warp of the outer Milky Way, known since 1957 \citep{kerr57}, has
been quantitatively determined for the first time by \citet{levine05}.
It can be described as a superposition of three and only three of the
lowest order vertical harmonics of a disk: a {\em dish-shaped} $m=0$,
a {\em integral-sign-shaped} $m=1$, and a {\em saddle-shaped harmonic}
$m=2$.  The lines of nodes for each are close to coincident and nearly
radial.  The amplitudes of each reaches 7--10\% of the radius of the
disk.  A number of possible warp producing mechanisms have been
suggested including long-lived eigenmodes, forcing by halo
triaxiality, persistent cold-gas accretion, and tidal excitation.  We
show here that the origin of this warp can be well-described as the
tidal interaction of the Magellanic Clouds with the disk and dark
matter halo of the Milky Way.  The interaction of the dark matter halo
with the disk and resonances between the orbit of the Clouds and the
disk account for the large amplitudes of the three harmonics and their
approximate shape and orientation.

\section{Observations}

\citet{levine05} found that a dynamical model for the warp must
satisfy six observational constraints: 1) the three lowest-order
harmonics: $m = 0, 1$ and $2$ are necessary and sufficient to describe
the global shape of the warp; higher-order global harmonics are
typically an order of magnitude or more weaker; 2) the $m = 1$ warp
has the largest amplitude everywhere in the outer disk; 3) the $m = 0$
and $m = 2$ warps are comparable in amplitude to the m = 1 warp, but
are smaller at all radii; 4) the $m = 1$ warp has a measurable
amplitude at the galactocentric radius of the Sun, $R_\circ$, but the
$m = 0$ and $m = 2$ warps begin near the edge of the stellar disk, at
$R = 2R_\circ$; 5) all three harmonics grow approximately linearly with
radius, reaching amplitudes of 1--2 kpc at about $R$ = 30 kpc; and 6)
the amplitudes of each of the harmonics reach 5--10\% of the radius
of the disk. The lines of maximum descent of the $m = 1$ and $m = 2$
warps are coincident within about 12 degrees in galactocentric azimuth
$\phi$ and show little evidence of precession. The lines are located
near $\phi = 90\degr$.

\section{Methodology}

We use the procedure described by \citet[W98]{weinberg98} to couple
the halo response to the tidal excitation theory presented by
\citet[HT]{hunter69}.  This assumes that the disk remains thin and
that gas dissipation is unimportant for the dynamics. HT derived an
expression for the bending of a thin disk by the Magellanic Clouds and
concluded that the direct excitation of the disk by the Clouds
produces a warp of only a few hundred parsecs, an order of magnitude
less than the amplitude of the observed warp.  W98 wedded the halo
excitation presented in \citet{weinberg89} to the HT approach,
allowing the disk to feel both the tidal field from the Clouds
directly as well as the force from the dark-matter halo wake excited
by the Clouds.  The assumption of linearity limits the predictions to
modest amplitudes. 

We use perturbation theory rather than N-body simulation because of
the intrinsic difficultly and subtlety in obtaining accurate multiple
time scale results from a simulation.  The excitation hierarchy of
satellite orbit$\rightarrow$halo wake$\rightarrow$disk bending modes
results in multiple interleaved time scales: the orbital periods of
the Clouds, the pattern speeds of the halo wakes, and the pattern
speeds of the bending modes. We also have multiple spatial scales. For
example, a simulated N-body disk must be capable of supporting the
bending modes.  In addition, particle simulations can degrade the
resonant dynamics as described by \citet{weinberg05}.  With such
difficulties, one should verify the dynamics of each component of the
mechanism before putting everything together in an N-body simulation,
although this is rarely done.

\section{Model}

The perturbation theory includes the force from an extended satellite.
However, the halo and disk excitation depend only on the lowest-order
hamonics while the spatial extent changes the higher-order harmonics,
and thus the extent plays little role.  Therefore, as long as the
Clouds remain bound, their masses may be added for our estimates.  We
use the disk profile from W98.  Our halo is an NFW profile with $c=15$
\citep{navarro97} and virial mass 20 times the disk mass.  The orbital
plane was computed as described as in W98. We used the radial velocity
and proper motion from \citet{kalli05} and the distance modulus from
\citet{freedman01} to derive a space velocity and computed the
resulting orbit in the spherical dark matter halo.  We adopted the LMC
mass from \citet{westerlund97} of $2\times10^{10}M_\sun$, although
there is considerable variance in the literature. The current position
of the LMC is shown as is the current state of the warp.  This orbit
will carry it toward the NGP.

Although our calculation assumes a collisionless and vertically thin
medium, it is more generally applicable for the following reasons.
Firstly, although the dispersion relations for a multicomponent and
collisionless media differ at small scales
\citep[e.g.][]{jog96,rafikov01}, the large-scale warp will be governed
by inertia and the gravitational restoring force, not by local
pressure.  In addition, the disk self gravity is dominated by the
inner stellar disk; the outer gas layer plays only a minor role in
establishing the modes.  Therefore, our collisionless results are
likely to be similar to a multicomponent calculation.  Secondly, the
vertical restoring force depends very weakly on the thickening as long
as the vertical degree of freedom does not couple to the bending (this
may be demonstrated by straightforward but tedious algebra).
Therefore, the modes will be largely unchanged by thickening over
short time scales.  However, the challenging problem of vertical
coupling is important and remains to be investigated thoroughly.

The warp is a very dynamic structure based on the temporal evolution
of the model.  This can be seen in the AVI file of the simulations
which can be found at {\tt http://www.astro.umass.edu/\textasciitilde
weinberg/lmc}. Also included at this site are comparisons of the $m =
0,1,2$ evolution.  Rather than a static structure that might be
expected for a warp in response to a triaxial halo, a warp that
results from the Magellanic Clouds is continuously changing shape
because of the varying amplitudes and phases of the various modes.
The image looks rather like a flag flapping in the breeze as the
Clouds completes an orbit of the Milky Way.

\section{Comparison with Observations}

\begin{figure*}
\plottwo{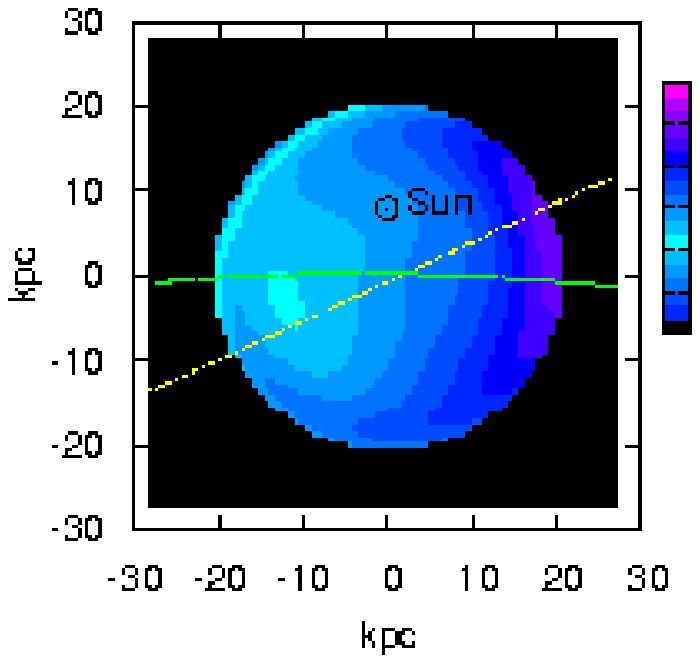}{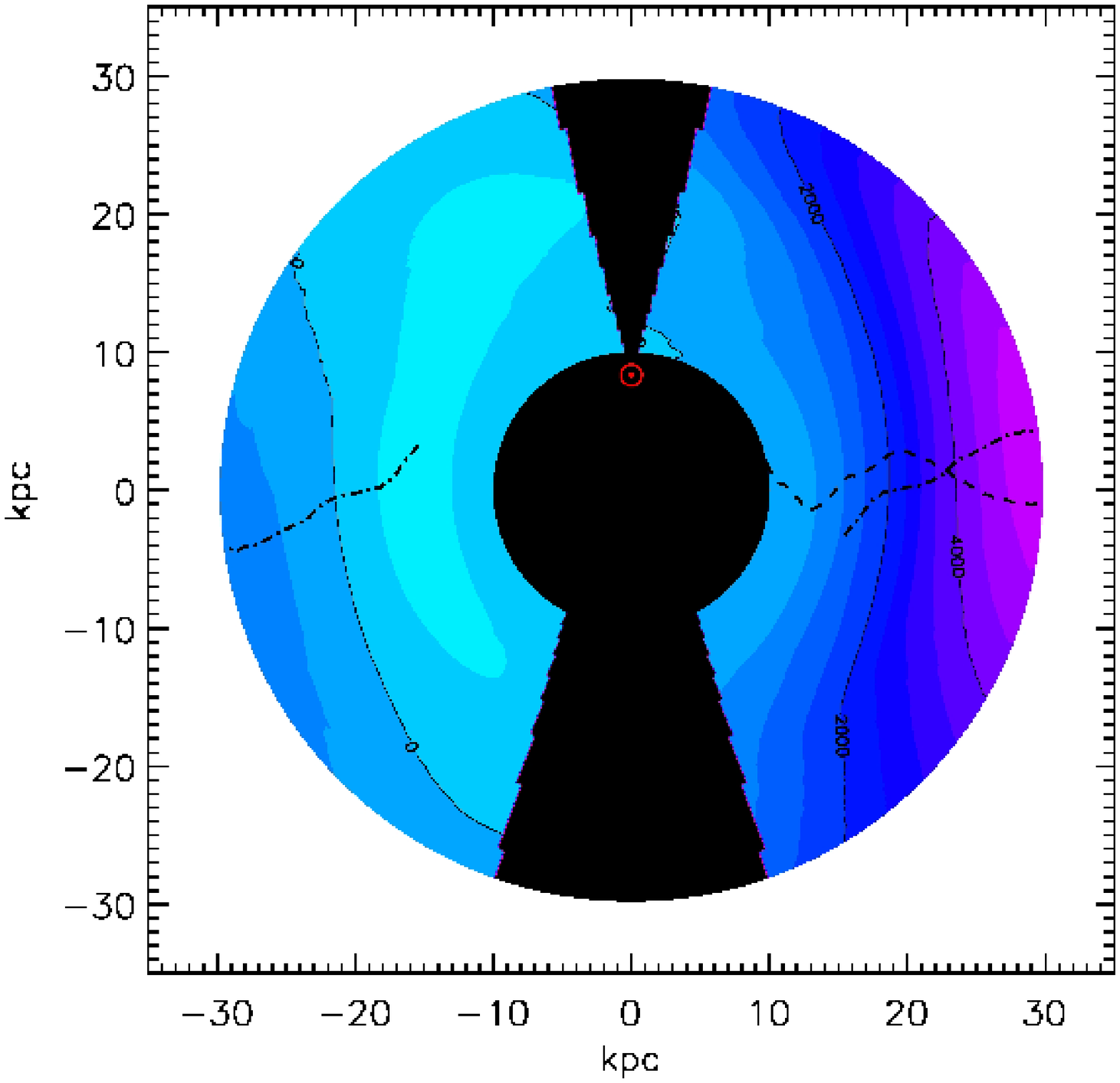}
\caption{Lines of maximum descent (perpendicular to the line of nodes)
for $m = 1,2$ from the model (dashed and dash-dot, resp.).  The
lines are superimposed on a contour plot of the deviation in $z$ of
the midplane of the disk from $b = 0\degr$. Compare with Fig. 11 in
\citet{levine05}.
\label{maxdescent}}
\end{figure*}

Figure \ref{maxdescent} shows the shape of the warp for both the model
and the data.  We plot only $m = 0,1,2$ from the data, ignoring the
weak but significant $m = 10, 15$ terms.  The overall agreement is
quite good, although there are some differences.  The model does not,
for example, have a minimum that is as extensive in Galactic azimuth
as the data.  Lines of maximum descent for $m = 1,2$ in both the
simulations and the HI data analysis are also shown in the figure.
They are separated by about 20$\degr$ in the model, a bit more than,
but close to the $12\degr$ in \citet{levine05}.  Both sets of lines
are oriented in the same sense: close to $\phi = 90\degr$.  Neither
set of lines shows evidence of significant variation with
galactocentric radius.

\begin{figure}
\plotone{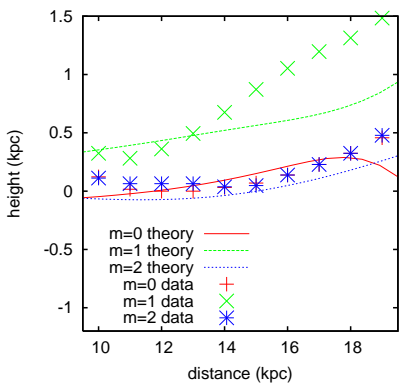}
\caption{The amplitudes of the $m = 0,1,2$ harmonics as a function of
Galactic radius from the simulations (curves) and from the HI Milky
Way data \citep[points][]{levine05}.
\label{amps}
}
\end{figure}

Figure \ref{amps} shows the amplitudes of the first three harmonics,
$m = 0,1,2$, in both the model and the data.  We increased the Clouds'
mass by 33\% for a better fit.  As in the data, the $m = 1$ in the
simulations is the strongest, and increases nearly linearly out to the
edge of the disk.  At large R, where the inertia is small, the linear
theory is expected to overpredict the warp \citep{tsuchiya02}.  The
simulations also show a weak response of $m = 0,2$ out to about 15
kpc, and then increasing nearly linearly, but with approximately the
same amplitude for both, providing a reasonable representation of the
data.
 
The amplitude ordering of these vertical harmonics has a natural
physical explanation.  The force from the halo wake and the satellite
may described by three-dimensional harmonics and these affect the warp
height as follows.  A spherically symmetric $m = 0$ halo distortion
does nothing to the warp. An $m = 1$ halo distortion will tend to
accelerate the disk in the vertical direction; the differential
acceleration of the disk results in a vertical $m = 0$ ``dishing''.
An $m = 2$ halo distortion will attract the disk upward and downward
in a reflection-symmetric way causing the classic integral-sign warp.
Higher-order symmetries may be deduced by similar geometric
considerations. The power in the halo excitation drops off as an
inverse power of the harmonic order and only the lowest order terms
have features well inside the satellite orbit.  Conversely, the
existence of these higher-order harmonics with a power-law drop off is
a natural consequence of this tidal theory and is consistent with the
data.
 
We fixed the disk and the halo mass inside of the virial radius while
adjusting the satellite orbit and halo concentration and found the
following trends.  First, the halo wake and its pattern speed are
determined by the halo concentration.  The disk bending modes have a
natural set of frequencies for a given halo.  These will be maximally
excited when forced by the halo wake at or near harmonics of this
natural frequency.  The ratio of $m=2$ amplitude to the $m=1$
amplitude is maximized for a halo concentration $c\approx10$.  For an
NFW profile, this puts $r_s\approx30$ kpc.  This is very close to the
$\Lambda$CDM estimates for the Milky Way concentration.  Secondly, the
orientation of the response depends on whether the nearest resonance
is larger or smaller than natural frequency.  Therefore changing the
satellite orbit, which changes the forcing frequency, affects both the
amplitude of the amplitude and orientation of the warp response.  In
short, the warp depends on a ``clockwork'' of frequency relationships
which depends on the satellite orbit, the dark-matter halo and the
disk.  A pericenter larger than the current 49 kpc estimate shifts the
position angle (PA) so that the warp peaks closer to PA=180 degrees.
Similarly, smaller pericenter increases the amplitude also shifts the
PA.  We conclude that our model ``prefers'' our current fiducial MC
model.  We are not claiming that our fiducial model is the most
probable amongst the distributions of allowed values but that a
plausible choice of parameters corresponds to many features of the
observed data.
  
\newpage

\section{Discussion}  

\subsection{Comparison with N-body Simulations}

The predictions of W98 were checked by several groups using N-body
simulations. \citet{garcia-ruiz02} used a hybrid N-body particle--ring
code and did not find the warp predicted in W98. They offer the
failure of the linear theory as the culprit but did not investigate a
variety of models. Similarly, \citet{mastro05} performed a simulation
including both the gaseous, stellar and dark components of the Large
Cloud and the Milky Way and remark that the effect on the Milky Way is
negligible.  Conversely, \citet{tsuchiya02}, using a hybrid code that
includes both a potential expansion and a tree code, obtained
amplitudes of $m = 0,1,2$ warps in the larger halo model that show
very good agreement with the observed amplitudes in the Milky Way. He
also shows that the amplitude of the warp depends strongly on the dark
halo mass model.

      It is difficult to reconcile these contradictory findings but
three possibilities obviously occur: 1) the linear theory does not
apply; 2) the simulations do not apply due to numerical difficulties;
and 3) the two simulations with null warps have chosen unlucky sets of
parameters.  Both \citet{garcia-ruiz02} and \citet{tsuchiya02} choose
methods that explicitly treat the multiple
scales. \citet{garcia-ruiz02} used tilted rings to represent the disk
and a tree code to represent the halo. Although the use of rings
limits the investigation to the $m = 1$ response, it should be
sufficiently sensitive to the halo excitation without strong particle
number issues.  \citet{tsuchiya02} used an expansion algorithm to
represent the halo gravitational field and a tree code to represent
the disk to better represent the multiple scales \citep[see][for
additional discussion of the dynamic range problems that motivate this
choice]{tsuchiya02}. We feel that the discrepancy between the results
of these two groups is most like Item (3): the Garc\'ia-Ruiz et
al. model is not particularly warp producing.  The good qualitative
correspondence between W98 and \citet{tsuchiya02} suggests that the
linear theory captures the underlying physics although is likely to
differ in detail. For example, the linear theory over-predicts the
height in the edge of the disk (Fig. \ref{amps}), and this motivates
our truncation of the predictions in the outer disk.  \citet{mastro05}
primarily emphasize the affect of the Milky Way tides on the Large
Cloud and do not tailor their approach to treat the multiple scale
problem per se. We suggest that their report of no disk warp results
from Items (2) and (3).

\subsection{Other Explanations}

Most warp theories depend on the bending response of the disk and this
response is strongly affected by the existence of a self-gravitating
dark matter halo.  The theories may be roughly grouped as follows: 1)
bending modes may be primordially excited and persistent
\citep[e.g.][]{sparke88}; 2) the response of the disk to the
non-axisymmetric shape of the halo; 3) tidal excitation (as we have
discussed here) and 4) the warp may be produced by the response of the
disk to cosmic infall \citep[e.g.][]{jiang99}. The topic was nicely
reviewed by \citet{binney91}. Subsequently, \citet{nelson95} argued
against long-lived modes.  The triaxiality of the Milky Way halo
remains uncertain.  \citet{helmi04} reports a prolate halo (q=1.25)
while \citet{johnston05} finds an oblate halo (q=0.8-0.9) using Sgr
dwarf constraints.  Such modest triaxiality seems unlikely to produce
the observed vertical $m=2$ feature. In addition, cosmic infall more
naturally produces tilted rings, an $m=1$ feature; the observed
vertical $m=2$ may require a conspiracy of several inflow directions.

\subsection{Relevance for Modified Gravity}

Although the success of our model favors the existence of a
dark-matter halo in Nature, many find it seductive to modify gravity
to produce the observed rotation velocities in the Galaxy without dark
matter \citep[modified Newtonian dynamics or
MOND][]{milgrom83a,milgrom83b,bekenstein84}. Might the tidal theory
also apply in MOND? It is beyond the scope of this letter to repeat
our calculations using MOND, it seems plausible that direct forcing of
the Galactic disk by the Clouds in MOND may provide warp amplitudes in
excess of the original predictions without a dark halo (HT).
Similarly, we would expect the disk modes and frequencies to be
qualitively similar, the excess restoring force of the halo being
produced by the MOND force.  However, MOND would have to (1) admit
bending modes with similar morphology to those in the Newtonian theory
and (2) these modes would have to conspire to frequencies that couple
them to the direct forcing by the Clouds in such a way that they
assumed the same orientation as in the ``clockwork'' described in \S5.
Because this clockwork and the $m=2$ to $m=1$ amplitude ratio depends
on the simultaneous halo and disk excitation, agreement with the
observations described in \S2 seems rather unlikely and thereby
disfavors MOND.  We encourage detailed predictions.

\subsection{Relevance for other Galactic Systems}  

A large fraction of other warped galaxies also show warps in their HI
layers and a significant fraction of these are asymmetric.  Warps in
these systems are generally analyzed with a program such at ROTCUR
\citep{begeman89} or one of its derivatives, but invariably are forced
to fit the $m = 1$ warp only.  This paper, and the work of
\citet{levine05} show that at least three harmonics ought to be fit to
the warps of external galaxies, especially those with asymmetric
warps, which can be caused by a superposition of these harmonics.  The
degree to which various harmonics are present in a warp can produce
important constraints on whether the warp is due to a satellite, a
triaxial halo, cold gas inflow, or some primordial excitation.  All
warps may not be alike.

\subsection{Summary}

We have demonstrated a plausible mechanism for the excitation of the
warp that explains all of its general features.  A number of warp
producing mechanisms have been explored elsewhere in addition to
tides.  It is possible that different mechanisms may act in different
galaxies and possibly in concert in a single galaxy. Nonetheless, our
simple model nicely reproduces the observed features in the Milky Way
HI gas layer.  The existence of massive companions, the Magellanic
Clouds, and the prediction from linear perturbation theory and at
least one corroborating N-body simulation suggests that a tidal
explanation is viable.  Our model depends on the gravitational
response of the halo and thereby suggests that the dark matter is not
an artifact of modifying the laws of gravity. Conversely, given a dark
halo, we argue that the tide from the Magellanic Clouds must be
affecting the Milky Way disk, and, given the quality of the agreement,
it seems to be the dominant mechanism.  This model then promises an
additional constraint on the distribution of dark matter. Although we
have emphasized the gas layer response beyond the stellar disk, the
effect of other satellite encounters such as the recent accretion of
Sagittarius dwarf may be detectable in future high-resolution surveys
and may help determine the properties of the inner halo.  Warp
observations are important because promise to reveal aspects of the
dark matter distribution that are otherwise observationally
inaccessible.  Further analysis of warps may provide more precise
constraints on the profile of dark matter in the Milky Way and nearby
external galaxies.

\acknowledgments{L.B. would like to acknowledge partial support from
NSF grant 02-28963. M.D.W. has been partially supported by NASA
NAG5-12038 and NSF 02-05969.  We thank Arend Sluis and the anonymous
referee for comments on the manuscript.}

\end{document}